\documentclass[fleqn,twoside]{article}
\usepackage{espcrc2}

\usepackage{graphicx}

\newcommand{\ds}{\partial\!\!\!\raisebox{2pt}[0pt][0pt]{$\scriptstyle/$}}

\usepackage{color}
\definecolor{rosso}{cmyk}{0,1,1,0.4}
\definecolor{rossos}{cmyk}{0,1,1,0.55}
\definecolor{rossoc}{cmyk}{0,0.5,1,0.2}
\definecolor{blu}{cmyk}{1,1,0,0.3}
\definecolor{blus}{cmyk}{1,1,0,0.6}
\definecolor{blucc}{cmyk}{1,0.4,0.2,0}
\definecolor{viola}{cmyk}{0,1,0,0.6}
\definecolor{viola2}{cmyk}{0,1,0.2,0.6}
\definecolor{verde}{cmyk}{0.92,0,0.59,0.25}
\definecolor{verdec}{cmyk}{0.92,0,0.59,0.15}
\definecolor{verdes}{cmyk}{0.92,0,0.59,0.4}
\font\tenrsfs=rsfs10 at 12pt
\font\sevenrsfs=rsfs7
\font\fiversfs=rsfs5
\newfam\rsfsfam
\textfont\rsfsfam=\tenrsfs
\scriptfont\rsfsfam=\sevenrsfs
\scriptscriptfont\rsfsfam=\fiversfs
\def\mathscr#1{{\fam\rsfsfam\relax#1}}

\newcommand{\fig}[1]{~\ref{fig:#1}}

\newcommand{\GeV}{\,{\rm GeV}}
\newcommand{\TeV}{\,{\rm TeV}}
\def\circa#1{\,\raise.3ex\hbox{$#1$\kern-.75em\lower1ex\hbox{$\sim$}}\,}

\newcommand{\NP}{Nucl. Phys.}

\newcommand{\PL}{Phys. Lett.}
\newcommand{\PR}{Phys. Rev.}
\newcommand{\beq}{\begin{equation}}
\newcommand{\eeq}{\end{equation}}

\def\circa#1{\,\raise.3ex\hbox{$#1$\kern-.75em\lower1ex\hbox{$\sim$}}\,}
\makeatletter

\def\art{\@ifnextchar[{\eart}{\oart}}
\def\eart[#1]#2#3#4#5#6{{\rm #2}, {\em #3 \rm #4} {\rm (#6) #5} ({\em #1})}
\def\hepart[#1]#2{{\rm #2, \em#1}}
\newcommand{\oart}[5]{{\rm #1}, {\em #2 \rm #3} {\rm (#5) #4}}

\newcounter{alphaequation}[equation]
\def\thealphaequation{\theequation\hbox to
0.6em{\hfil\alph{alphaequation}\hfil}}
\def\eqnsystem#1{
\def\@eqnnum{{\rm (\thealphaequation)}}
\def\@@eqncr{\let\@tempa\relax \ifcase\@eqcnt \def\@tempa{& & &} \or
  \def\@tempa{& &}\or \def\@tempa{&}\fi\@tempa
  \if@eqnsw\@eqnnum\refstepcounter{alphaequation}\fi
\global\@eqnswtrue\global\@eqcnt=0\cr}
\refstepcounter{equation} \let\@currentlabel\theequation \def\@tempb{#1}
\ifx\@tempb\empty\else\label{#1}\fi
\refstepcounter{alphaequation}
\let\@currentlabel\thealphaequation
\global\@eqnswtrue\global\@eqcnt=0 \tabskip\@centering\let\\=\@eqncr
$$\halign to \displaywidth\bgroup \@eqnsel\hskip\@centering
$\displaystyle\tabskip\z@{##}$&\global\@eqcnt\@ne
\hskip2\arraycolsep\hfil${##}$\hfil& \global\@eqcnt\tw@\hskip2\arraycolsep
$\displaystyle\tabskip\z@{##}$\hfil
\tabskip\@centering&\llap{##}\tabskip\z@\cr}
\def\endeqnsystem{\@@eqncr\egroup$$\global\@ignoretrue} \makeatother

\newcommand{\MeV}{\,\hbox{\rm MeV}}
\newcommand{\eV}{\,\hbox{\rm eV}}

\newcommand{\AmS}{{\protect\the\textfont2
  A\kern-.1667em\lower.5ex\hbox{M}\kern-.125emS}}

\title{Searches for sterile neutrinos (and other light particles)}

\author{Alessandro Strumia\address[MCSD]{Dipartimento di Fisica
dell'Universit\`a di Pisa and INFN, Italia}}
            
\begin{document}

\begin{abstract}
Future neutrino and cosmological experiments will 
perform powerful searches for new light particles.
After a general introduction we discuss how
a new light fermion (`sterile neutrino') could manifest.
\vspace{1pc}
\end{abstract}

\maketitle

 In the past years cosmology and neutrino experiments discovered new physics,
 which in both cases  presently looks `standard': 
 oscillations of 3 massive neutrinos,
 and $\Lambda$CDM with scale-free 
  primordial fluctuations.
  In the next years new experiments will test such `standard models'
 and will try to detect the unseen effects that they suggest, such as
 $\theta_{13}$, 
 $\theta_{23}-\pi/4$,\ldots;
 tensor fluctuations, $n-1$,...
 Discovering something unexpected would be more interesting:
 to which kind of new physics these future experiments will be particularly sensitive?
 
 In both cases the answer seems: new light particles.

 \section{New light particles}
Any new light particle would be an important discovery, because
its lightness would presumably be due to some deep reason
(as is the case for $\gamma,\nu$ and gravitons).

 Experimentally, the possible existence of new light particles
can be tested by studying how known light particles
($\gamma$, $\nu$ and gravitons) behave in different situations.
Extreme conditions can be tested in cosmology and astrophysics.

Concrete examples of possible manifestations of new light particles
are axion/photon interactions and 
infra-red modifications of gravity.
We here discuss what can be done studying neutrinos.

\smallskip
 
Neutrinos are not the best tool for discovering new
 heavy particles, because at high energy $\nu$ interact 
 like left-handed charged leptons, and it is easier to work with $e$.
 Furthermore theorists suggested (based on quite general effective-Lagrangian arguments)
that new physics could manifest at low energy making neutrinos massive, 
but without affecting other neutrino properties.
Experiments are now confirming this view, by indicating that the new physics behind
solar~\cite{sun} and atmospheric~\cite{atm} anomalies is neutrino oscillations.

\begin{figure*}[t]\vspace{-4mm}
$$\includegraphics[width=6cm]{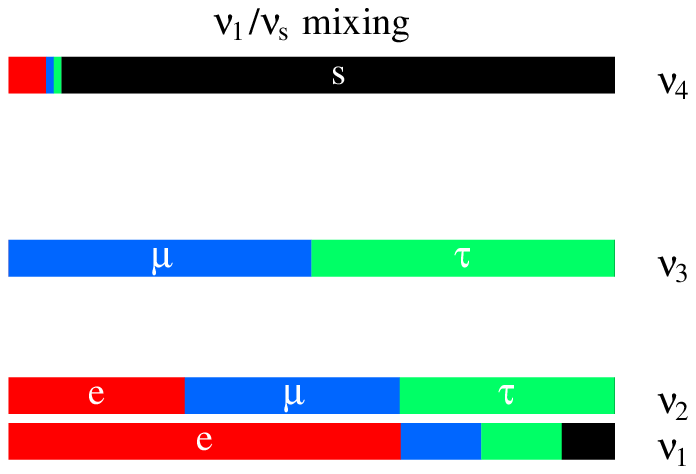}\qquad\qquad
\includegraphics[width=6cm]{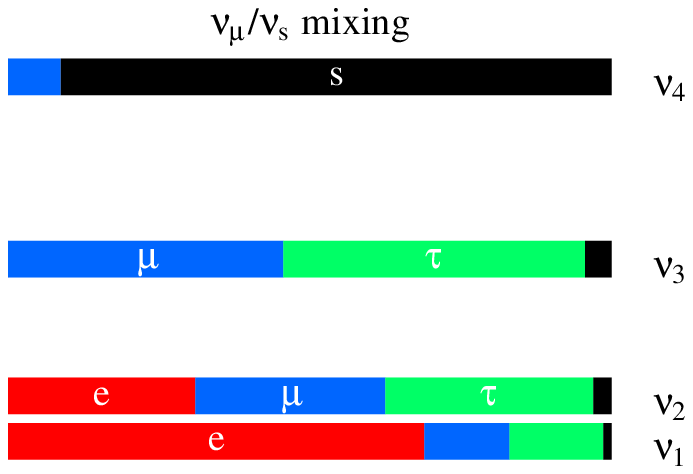}$$\vspace{-14mm}
\caption{\label{fig:spettrias}
Basic kinds of four neutrino mass spectra. 
Left: sterile mixing with a mass eigenstate ($\nu_1$ in the picture).
Right: sterile mixing with a flavour eigenstate ($\nu_\mu$ in the picture).}
\end{figure*}

\begin{figure*}[p]
$$\includegraphics[width=6cm,height=6cm]{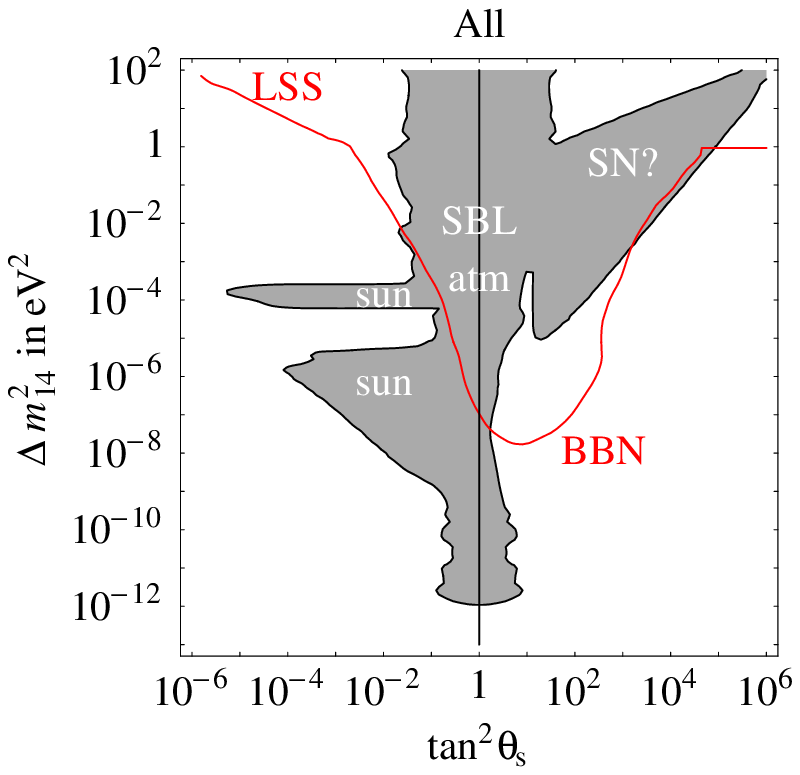}\qquad\qquad
\includegraphics[width=6cm,height=6cm]{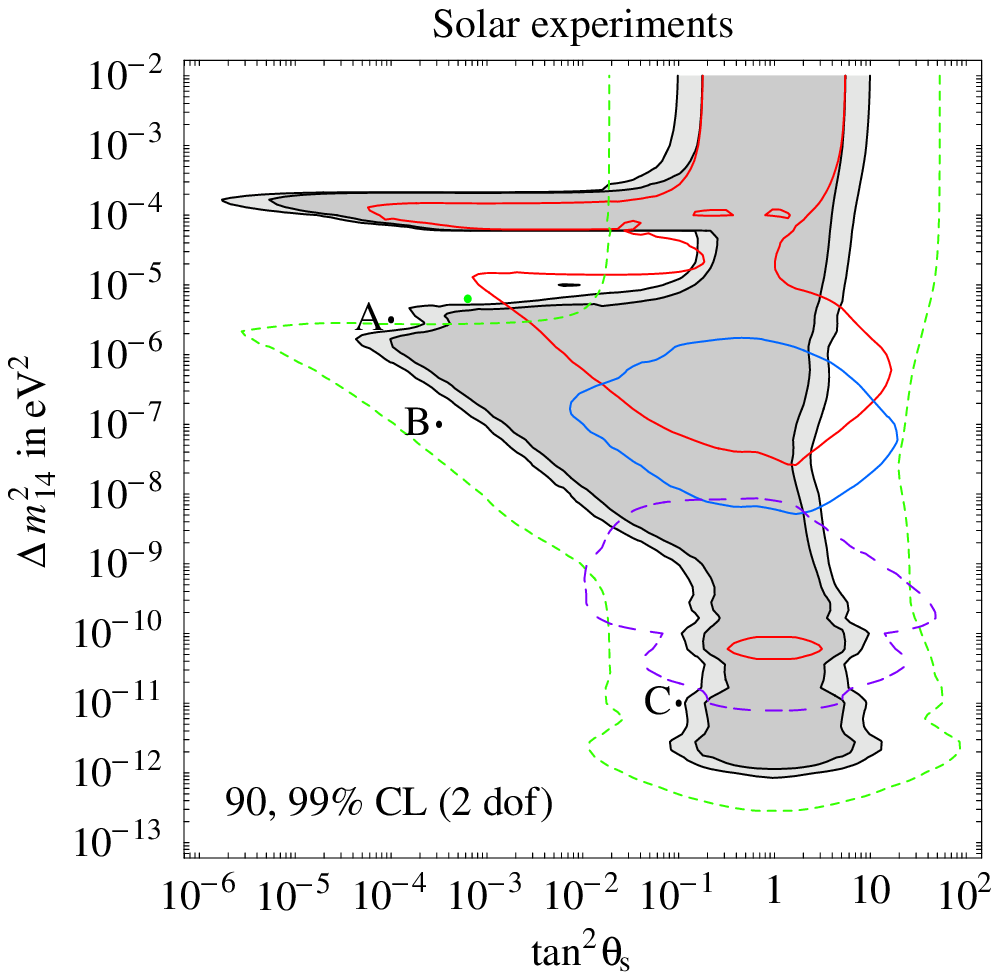}$$
$$\includegraphics[width=6cm,height=6cm]{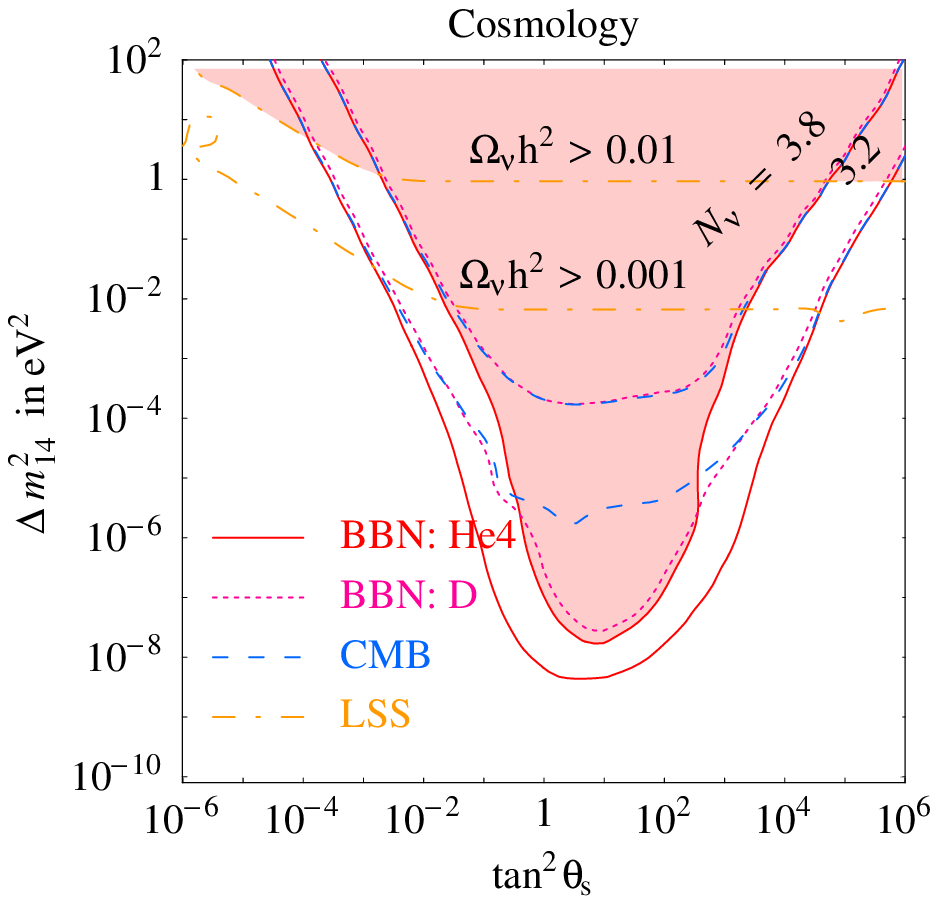}\qquad\qquad
\includegraphics[width=6cm,height=6cm]{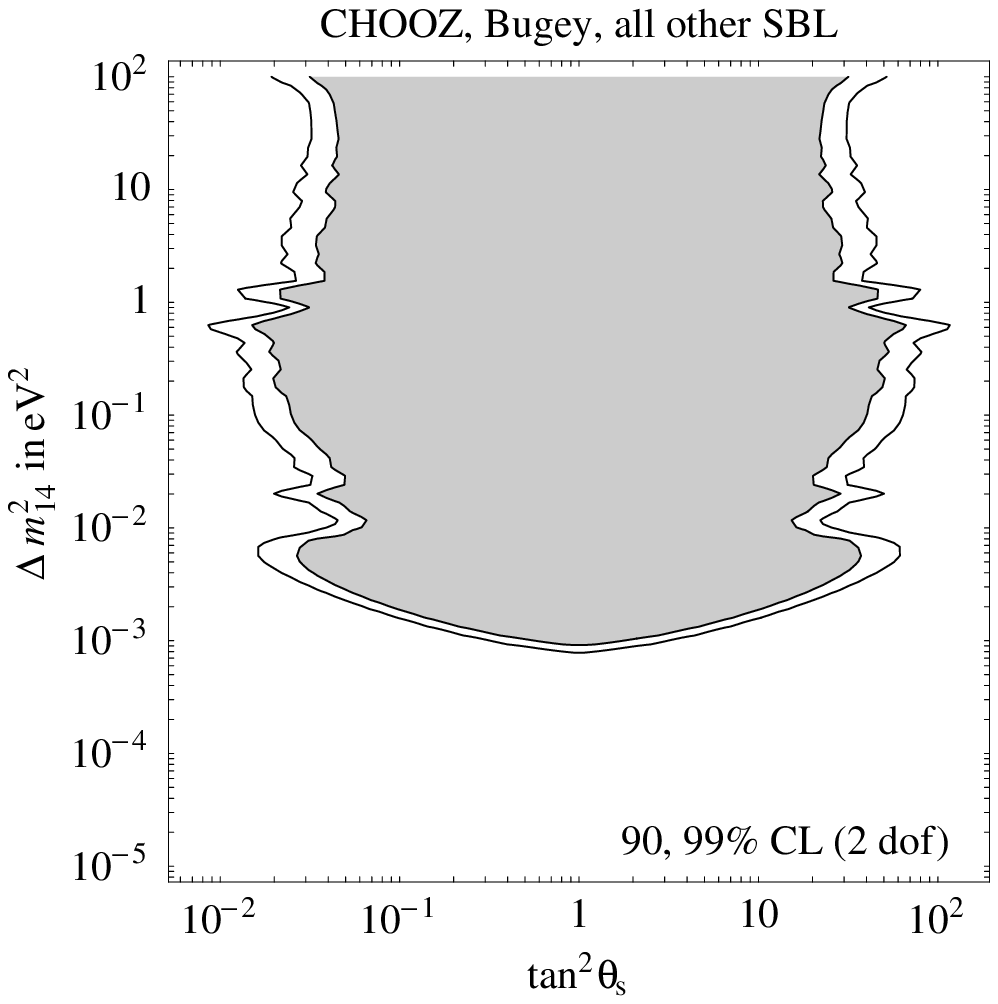}$$
$$
\includegraphics[width=6cm,height=6cm]{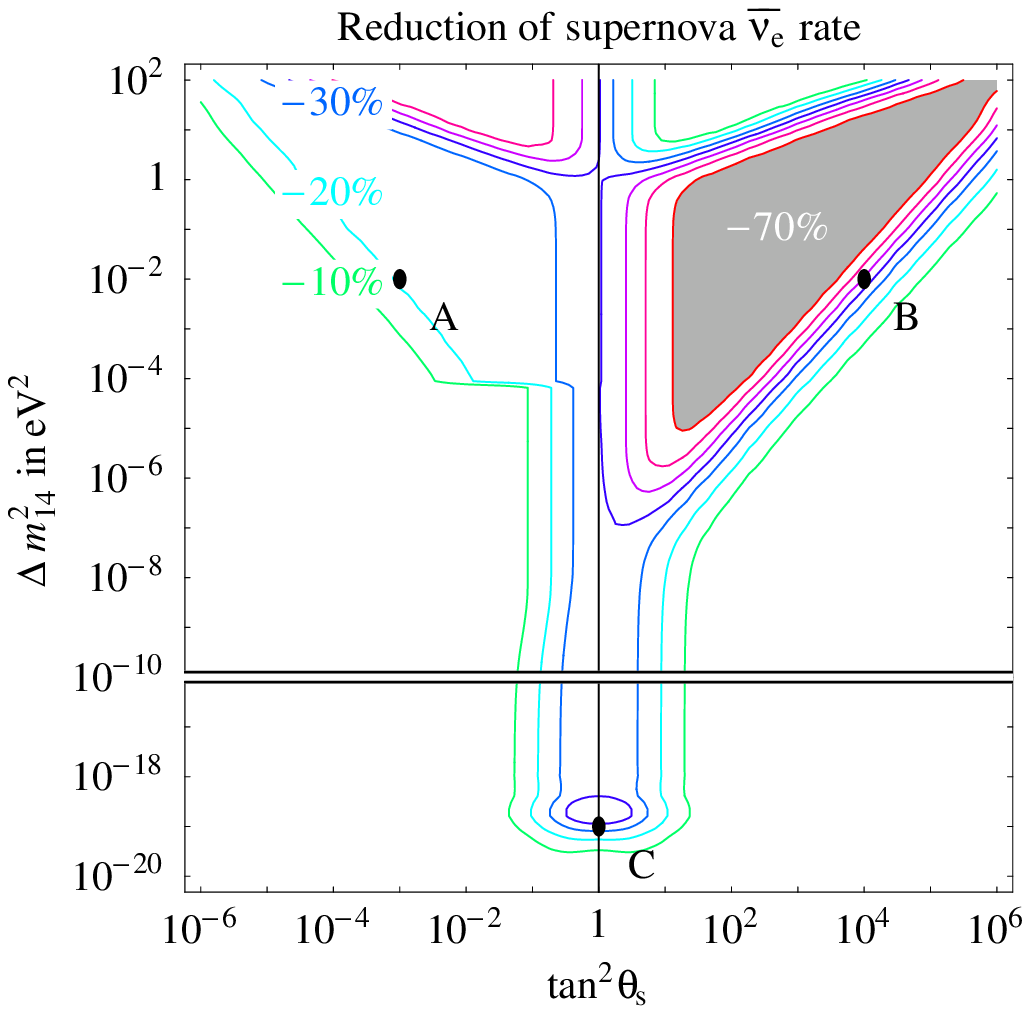}\qquad\qquad
\includegraphics[width=6cm,height=6cm]{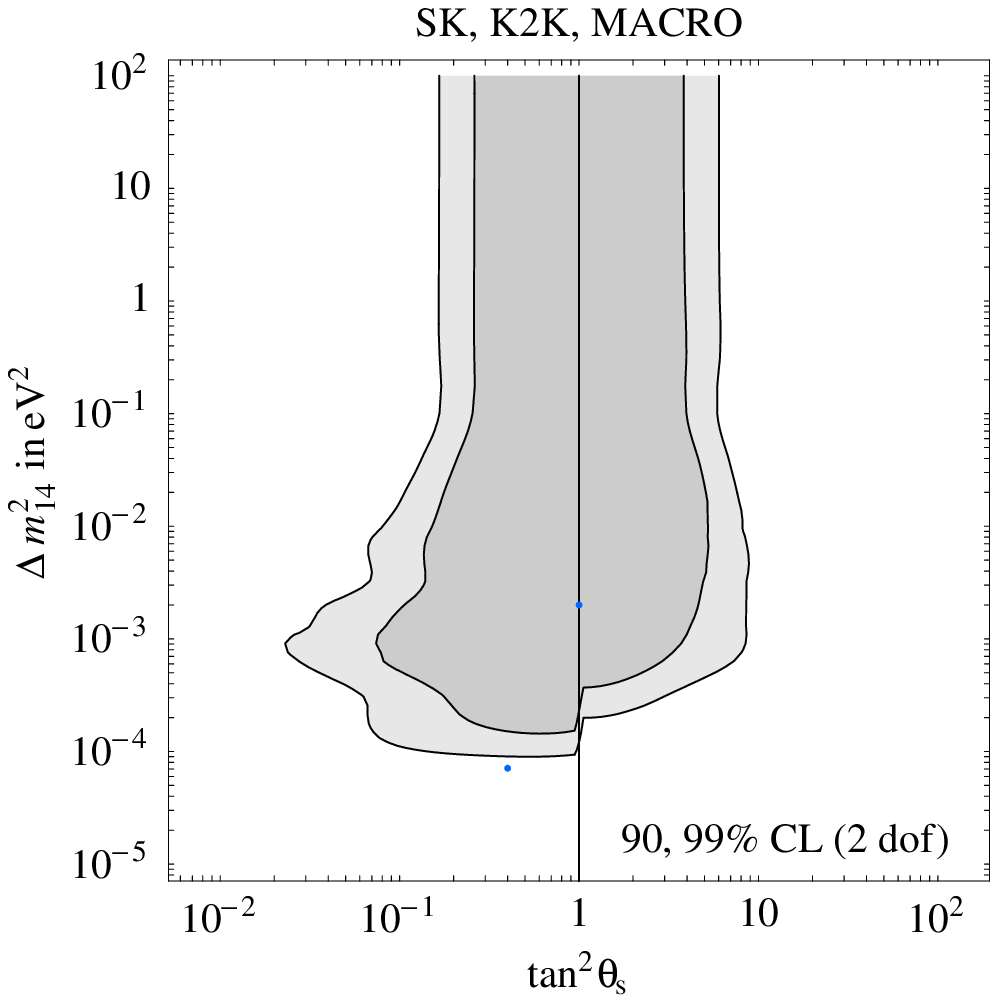}$$
\caption{\label{fig:S1}Fig.\fig{S1}a (from~\cite{CMSV})
summarizes the present status of $\nu_{\rm s}/\nu_1$ mixing,
showing regions already excluded or disfavored by a variety probes.
Each probe has a dedicated figure.
Solar neutrinos in fig.\fig{S1}b.
Cosmology in fig.\fig{S1}c.
Short-baseline experiments in fig.\fig{S1}d.
Supernova neutrinos in fig.\fig{S1}e.
Atmospheric experiments in fig.\fig{S1}f.
These figures indicate what can be probed by future experiments.}
\end{figure*}

\begin{figure*}\vspace{-4mm}
$$\includegraphics[width=16cm]{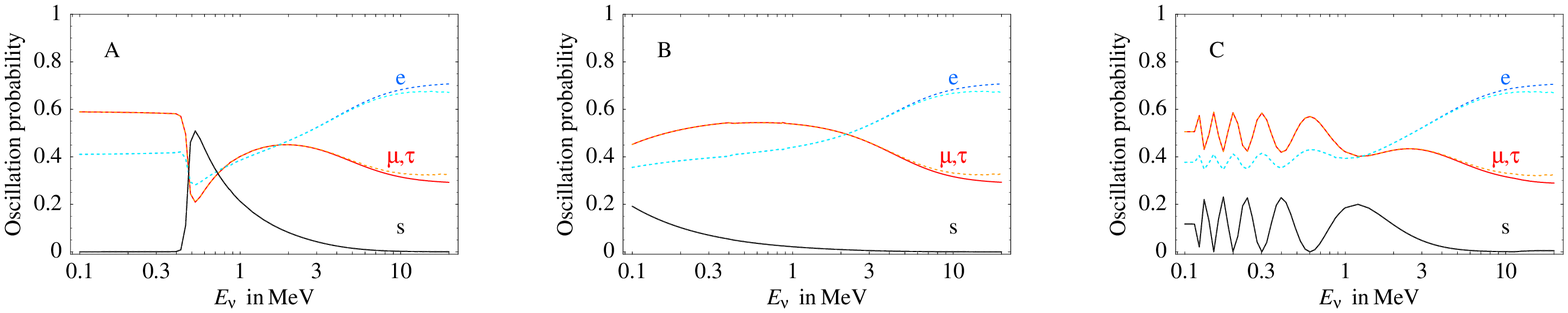}$$
\vspace{-1.3cm}
\caption{Oscillation probabilities $P(\nu_e\to \nu_{e,\mu,\tau,{\rm s}})$
of solar neutrinos in presence of $\nu_1/\nu_{\rm s}$ oscillations
at the benchmark points A, B, C of fig.\fig{S1}b.
Sterile effects are concentrated at low energies
(at intermediate energies in case~\cite{Smirnov}).}
\end{figure*}
\medskip

These theoretical arguments assume that no new light particle exist.
In the opposite case neutrinos can easily reveal other non-standard properties.
Interactions of neutrinos with a new light particle of mass $m$
depend on its spin:
a light scalar $\varphi$ can have Yukawa couplings $g \,\nu\nu \varphi$
(possibly of special type, if $\varphi$ were light because it is a Goldstone boson);
a light fermion $\nu_{\rm s}$ can have mass mixing with neutrinos
$m\,\nu\nu_{\rm s}$
(Goldstinos could have something more complicated);
a light vector can have new gauge couplings $g\,\bar\nu A\hspace{-1.3ex}/\, \nu$.

These couplings can give anomalous MSW effects
(of the form $g^2/m^2$ even if $m<E_\nu$),
make neutrinos decay, or prevent them to freely move in the early universe.
All this can be tested.
In some range of $m,g$ these effects are detectable compatibly with constraints from
energy losses in stars, supernov\ae, and the universe.

\smallskip

Light fermions have some appealing features.
Small fermion masses are easily stable under quantum corrections.
The SM contains fermions ($e,\nu, u,d,\ldots$)
variously charged under electric, weak, strong interactions:
fermions neutral under all these interactions
might exist without giving any observable effect in collider experiments.
A new light neutral fermion can naturally interact only with other
neutral light fermions: i.e.\ only via a mass mixing with neutrinos.

Adding an extra light fermion (a `sterile neutrino' $\nu_{\rm s}$)
is the simplest extension of the massive neutrino scenario which is emerging from data.
Indeed sterile neutrinos have been used as tentative interpretations of many anomalies
(solar, atmospheric, LSND, NuTeV, pulsar kicks, r-nucleosynthesis, Karmen,
low Chlorine rate, upturn in solar spectrum, solar time dependence, warm dark matter,
reionization, galactic $\bar{e}$, lower Gallium rates,\ldots).
So far this idea remained sterile.

\section{Searching for sterile neutrinos}
Adding one extra sterile neutrino,
the $4\times4$ neutrino mixing matrix is described by a few extra 
parameters: it is convenient to choose one active/sterile mixing angle $\theta_{\rm s}$
and a versor $\vec{n}$ that specifies which combination of active neutrinos $ \vec{n}\cdot\vec{\nu}$
mixes with $\nu_{\rm s}$. 
We consider a few representative cases:
\begin{itemize}

\item {\em Mixing with a mass eigenstate}  (fig.\fig{spettrias}a):
$\vec{n}\cdot\vec{\nu}=\nu_i$ ($i=1$ or 2 or 3).\
The sterile neutrino oscillates into a neutrino of mixed flavour
at a single $\Delta m^2$, which can be arbitrarily small.

\item  {\em Mixing with a flavour eigenstate}  (fig.\fig{spettrias}b):
$\vec{n}\cdot\vec{\nu}=\nu_\ell$ ($\ell =e$ or $\mu$ or $\tau$).
The sterile neutrino mixes with a well defined flavour oscillating 
at 3 different $\Delta m^2$
(two of them are known and equal to $\Delta m^2_{\rm sun,atm}$).

\end{itemize}
We emphasize that the usual `bounds on the sterile admixture' in solar or atmospheric
oscillations correspond to very special limiting cases,
which badly represent the general situation.

For example, a solar $\nu_e$ is usually assumed to oscillate into an {\em energy-independent}
combination of active and sterile neutrinos,
$$\nu_e\to  \eta~\nu_{\rm s} + \sqrt{1-\eta^2} ~\nu_{\mu,\tau}.$$
Measurements  of higher energy solar neutrinos performed by SNO and SK
presently
imply the dominant constraint,
$\eta = 0\pm0.1$.
This energy-independent sterile effect can be obtained 
from a heavy $\nu_{\rm s}$ which mixes with $\nu_\mu$, as
depicted in fig.\fig{spettrias}b.
However the generic situation can be even qualitatively different:
LMA oscillations provide a mechanism which can hide sterile effects 
in the range of neutrino energies explored by SNO.
Let us recall how LMA behaves in absence of sterile neutrinos.
There is a critical energy $E_* \sim {\Delta m^2_{\rm sun}}/{G_{\rm F} N_e^{\rm sun}} \sim\hbox{few MeV}$.
At $E_\nu\gg E_*$ solar matter effects are dominant and adiabatic:
 the sun emits only $\nu_2$ so that $P_{ee} \simeq \sin^2\theta_{\rm sun}$.
At $E_\nu\ll E_*$ matter effects are negligible and
 the sun emits mostly $\nu_1$, giving $P_{ee} = 1 - \frac12 \sin^2 2\theta_{\rm sun}$.

Therefore a sterile neutrino that either mixes only with $\nu_1$,
or that experiences a MSW resonance with $\nu_1$ can affect solar neutrinos
{\em almost only at sub-MeV energies}, so far explored only by Gallium experiments.
This argument shows that  searches for sterile neutrinos are one 
concrete motivation for doing
sub-MeV solar neutrino experiments.
(If instead active/active oscillations were the end of the story
sub-MeV experiments would not significantly improve our knowledge of oscillation parameters~\cite{subMeV}).


\section{Status of $\nu_{\rm s}/\nu_1$ mixing}
Fig.\fig{S1}a shows the present status of $\nu_{\rm s}/\nu_1$ mixing
(which is pictorially illustrated in fig.\fig{spettrias}a).
The background of atmospheric and solar oscillations and its uncertainties
have been fully included, assuming $\theta_{13}=0$ and `normal' mass
hierarchy of active neutrinos.
The region with $\tan\theta_{\rm s}<1$ corresponds to $\nu_{\rm s}$ heavier than $\nu_1$
and has so far been mainly probed by solar experiments.
The region with $\tan\theta_{\rm s}>1$ corresponds to $\nu_{\rm s}$ lighter than $\nu_1$
and has so far possibly probed by SN1987A.
Cosmology can be non standard on both sides.
See~\cite{CMSV} for a  precise description.
We now examine each probe in greater detail discussing how it can be improved.

\subsection*{Solar neutrinos}
Solar neutrinos are a powerful discovery tool, because in their 
trip from the sun to the earth they experience a variety
of different processes.

\begin{figure}[t]\vspace{-3mm}
$$\includegraphics[width=7cm]{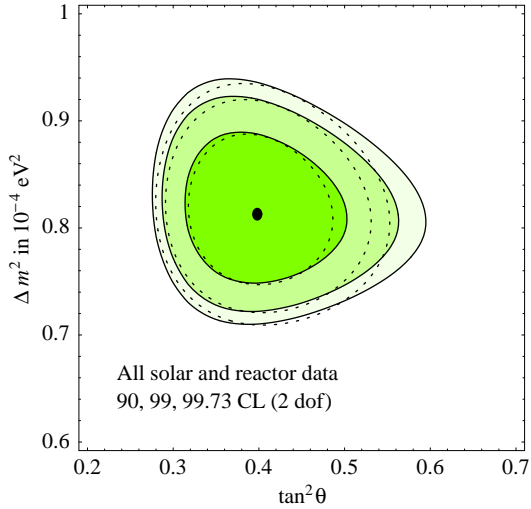}$$
\vspace{-13mm}
\caption{\label{fig:sun}Global active-only fit of solar oscillation parameters (continuous lines),
compared to its Gaussian approximation (dotted lines).}
\end{figure}

Fig.\fig{S1}b shows the region of $\nu_1/\nu_{\rm s}$ oscillation parameters
already tested (and excluded) according to the analysis of~\cite{CMSV}.\footnote{Fitting parameters
in Gaussian approximation simplifies computations, allowing to extract the values of
 $n$ parameters from
 $(1+n)(2+n)/2$ sampling points rather than the usual $(10\div 100)^n$.
This approximation is used for active oscillations parameters~\cite{CMSV}.
Fig.\fig{sun} shows how accurately it reproduces the up-to-date standard
 global fit of solar and reactor data.
}
It looks like a MSW triangle, 
strongly distorted by active/active LMA mixing.
Gallium experiments play a crucial r\^ole in the global fit.
The green dotted line shows how the probed region would extend if we could
test a $pp$ rate $2\%$ different from its LMA value.

Part of this region can be soon tested by Borexino~\cite{Borexino}, 
sensitive  mostly to the intermediate-energy Beryllium neutrinos.
Inside the blue continuous (dashed) lines their rate depends on time,
giving day/night (seasonal) asymmetries larger than $2\%$.
LMA predicts negligibly small asymmetries.
Sterile neutrino can also give anomalous earth-matter effects at high-energies.
Inside the continuous red line the day/night asymmetry at higher energies
(in the energy range measurable by a future Mton W\v{C} detector~\cite{Mton})
differs from LMA by more than $0.005$.
We normalize asymmetries as $(\max-\min)/{\rm mean}$.

\subsection*{Cosmology}
We recall that active/sterile oscillations with
$\Delta m^2 \circa{>}10^{-5}\eV^2$ can thermalize $\nu_{\rm s}$ before
neutrino decoupling, which occurs at a temperature $T\sim \MeV$.
Oscillations with smaller $\Delta m^2$ can only produce $\nu_{\rm s}$ 
depleting $\nu_{e,\mu,\tau}$, such that the total number of neutrinos remains constant.

We can test compatibility with standard cosmology, which however is not an established theory.
Non-standard cosmology, such as large neutrino asymmetries, 
might invalidate all the following discussion.
It is therefore important that we can measure at least 5 different observables that
test different aspects of neutrino behavior in different stages of the universe evolution.
For ease of presentation we re-express observables in terms of 
effective number of neutrinos `$N_\nu$'.
At the moment none of them is sensitive enough to distinguish $N_\nu = 3$ (standard case) from
$N_\nu=4$ (one thermalized extra light fermion) nor from $N_\nu = 3+ 4/7$ (one thermalized  extra light scalar).

\bigskip

\noindent{\bf 1.} The Helium-4 abundancy.  Presently it gives the most accurate determination
of a cosmological number of neutrinos (see ref.s in~\cite{CMSV}): 
$$N_\nu^{^4{\rm He}} = ~\stackrel{?}{2.7}~\pm\stackrel{?}{0.7},$$
which does not exclude a fourth thermalized sterile neutrino,
especially in view of the fact that both the central value and the error are controversial.

What is the physical meaning of $N_\nu^{^4{\rm He}}$?
It is related to the number of neutrinos thermalized at $T\sim \MeV$,
but differs from it in one important way.
The Helium-4 abundancy is also directly affected by scatterings involving $\nu_e,\bar\nu_e$
(such as $\nu_e n\leftrightarrow e p$)
 so that it is sensitive to their eventual depletion (see ref.s in~\cite{CMSV}).
 Due to this reason the sensitivity of the Helium-4 abundancy to sterile
 oscillations
 extends down to $\Delta m^2_{41}\sim 10^{-8}\eV^2$,
 as shown in fig.\fig{S1}c.
 
 
\bigskip

\noindent{\bf 2.} The Deuterium abundancy, $N_\nu^{\rm D}$,
is less sensitive to $\nu_e$ depletion than $N_\nu^{^4{\rm He}}$
and today has a larger error  (see ref.s in~\cite{CMSV}):
$$ N_\nu^{\rm D}  \approx 3\pm 2.$$ 
However it might be easier to improve on it.

\bigskip

\noindent{\bf 3.} The total energy density 
at recombination ($T\sim  \eV$) is blind to neutrino flavour and
today has a significant error, something like
$$ N_\nu^{\rm CMB}  \approx 3\pm 2$$
(see~\cite{NnuCMB} for more precise global fits).
Future measurements of Cosmic Microwave Background
(CMB) anisotropies can reduce its error down to $\pm0.2$ or better.

\bigskip

\noindent{\bf 4.} Neutrinos that freely move in the early universe make
dark and normal matter less clustered~\cite{Lyth}.
Neutrino masses suppress this effect on large scales
because non relativistic particles move slower (today $T\sim\hbox{meV}$).
Studies of galaxy clustering data presently imply~\cite{WMAP}
$$\Omega_\nu h^2=\frac{ \sum_i m_{ \nu_i} 
 N_{\nu_i}^{\rm CMB}}{93.5\eV}\circa{<} 0.01.$$
 The precise value depends on arbitrary details of the analysis.
 In absence of sterile neutrinos and assuming that active neutrinos
 have the `standard' abundancy,  $N_{\nu_i}^{\rm CMB}=1$, 
the above constraint implies $\sum m_{\nu_i} \circa{<} \eV$~\cite{WMAP}.
It also constrains an additional population $ N_{\nu_{\rm s}}$ of sterile neutrinos.
Fig.\fig{S1}c shows that it is a significant constraint.\footnote{See~\cite{CMSV} for 
cautionary remarks:
the precise constraint on
relatively heavy sterile neutrinos with relatively small abundancy has never been precisely extracted from data.}

\bigskip

\noindent{\bf 5.} Acoustic oscillations of CMB
are influenced by the existence of particles with negligible couplings, neutrinos.
New light particles can provide new neutrino interactions
that prevent them to freely move.
A neutrino mean free path smaller than the horizon size at recombination
would shift the `positions' of the CMB peaks,
 roughly by~\cite{NoFreeStream}
$$\Delta\ell = -8 (N_\nu^{\rm FS}-3)$$
if only $N_\nu^{\rm FS}$ neutrinos move freely.
Sterile neutrinos with significant self-interactions can realize this effect~\cite{NoFreeStream}.
This is not the case in the minimal scenario considered in fig.s\fig{S1},
where this effect is absent.

\bigskip

Direct detection of CMB neutrinos could become someday possible.
The expected flux  is orders of magnitude too small for present experiments.
Possible extra interactions of neutrinos would modify their clustering properties.
Can this result into a significant enough local overdensity?
The answer seems no.

\subsection*{Astrophysics}
So far we only detected neutrinos (presumably $\bar\nu_e$) from SN1987A, so 
we start discussing supernov\ae{}.
Fig.\fig{S1}e shows the reduction in the total $\bar\nu_e$ rate caused
by oscillations into sterile neutrinos.
Its behavior is dictated by MSW resonances: the main effect is at $\tan\theta_{\rm s}>1$
(rather than at $\tan\theta_{\rm s}<1$, as in the case of solar $\nu_e$),
because matter effects change sign going from neutrinos to anti-neutrinos.
A SN produces all active neutrinos: this gives the milder resonant effect at $\tan\theta_{\rm s}<1$.

The additional features at  $\Delta m^2_{14}\circa{>}\eV^2$ are present
because at such large values of $\Delta m^2_{14}$ new adiabatic resonances occur in 
the `deleptonized' interior region of the SN, where the electron matter potential flips sign.

The maximal suppression roughly corresponds to a $\cos^2\theta_{\rm sun}\sim 70\%$ deficit~\cite{SN3+1,CMSV}:
a $100\%$ deficit is no longer obtained because, due to active/active mixing,
$\bar\nu_e$ is no longer a mass eigenstate.

It is not clear if a $70\%$ deficit is already excluded by SN1987A observations,
nor if it can be tested in the future.
The problem is the uncertainty in the unoscillated flux.
Maybe it will only possible to probe spectral distortions, produced by sterile neutrinos
only around the borders of the MSW regions (e.g.\ at points A,B of fig.\fig{S1}e)
and in the vacuum oscillation region, around point C.
As in the solar case, sterile oscillations can also modify earth matter effects.

\bigskip

In a near future SK could detect relic $\bar\nu_e$ emitted by
past core-collapse supernov\ae~\cite{GdSK}.
Sterile neutrinos can suppress this flux (which however
cannot be precisely computed)
but can hardly give significant spectral distortions.

\bigskip

Various experiments are trying to detect neutrinos
emitted by other astrophysical sources~\cite{NuTelescopi}.
Such neutrinos would likely originate from $\pi^\pm$ decay chains
in regions with small matter density:
$$\pi ^-\to \bar\nu_\mu \mu^- \to  \bar\nu_\mu \nu_\mu\bar\nu_e e^-.$$
Therefore one expects initial total fluxes with the usual flavour ratio
$$N_e:N_\mu:N_\tau \approx 1:2:0$$
(different flavours have different energy spectra).
Active/active oscillations convert these fluxes into something like $$N_e:N_\mu:N_\tau \approx 1:1:1.$$
It is hard for sterile neutrinos to modify the flavour ratio $N_\mu:N_\tau$.
It is hard to measure the flavour ratio $N_e:N_{\mu,\tau}$, which can be
modified by sterile neutrinos (see e.g.~\cite{UHE,CMSV}).

\subsection*{Terrestrial experiments}
Sterile neutrinos can give disappearance of active neutrinos, but also
flavour conversions among them.

Fig.\fig{S1}d shows constraints on $\nu_1/\nu_{\rm s}$ mixing parameters
from terrestrial experiments not sensitive to the atmospheric anomaly.
Reactor experiments play the main r\^ole.

Fig.\fig{S1}f shows constraints 
from SK, K2K and {\sc Macro} data~\cite{Galla,SK2K} according the the analysis of~\cite{CMSV}.
This  second  group of experiments is more sensitive 
to smaller values of $\Delta m^2_{14}\circa{>}10^{-4}\eV^2$,
while the first group is more sensitive to small $\theta_{\rm s}$.
Therefore 
more precise reactor experiments with short ($\sim \hbox{km}$)
baseline~\cite{theta13} will give extended searches.

Constraints on sterile neutrinos from atmospheric data are usually studied
in a specific limiting case, namely assuming 
$$\nu_\mu\to  \eta~\nu_{\rm s} + \sqrt{1-\eta^2} ~\nu_{\tau}$$
oscillations at a single $\Delta m^2=\Delta m^2_{\rm atm}$.
Matter effects become large at higher energies, $E_\nu\gg \GeV$,
suppressing sterile mixing and thereby
allowing to discriminate the two channels.\footnote{At lower energies NC-enriched and $\tau$-enriched samples of SK data directly discriminate $\nu_\mu\to\nu_\tau$ from $\nu_\mu \to \nu_{\rm s}$.
But presently these samples are statistically less significant.}

Sterile neutrinos with  $\Delta m^2_{14}$ comparable to $\Delta m^2_{\rm atm}$
manifest also in new ways, giving
oscillations at multiple frequencies  (distorted by matter effects)
at $E_\nu\sim\GeV$.
Averaged oscillations are clearly seen in SK data, 
which however cannot see a clear oscillation pattern,
because SK cannot well reconstruct the neutrino energy.
Dedicated atmospheric detectors~\cite{Galla}
could test if the atmospheric oscillation pattern is standard.

A few $\nu_\mu$ beam experiments will soon precisely study
oscillations around the restricted range of $\Delta m^2$
suggested by atmospheric experiments~\cite{short}.
It is easy to compute possible effects of sterile neutrinos
and to precisely analyze the sensitivity of each experiment,
but it is difficult to make useful general statements.

\bigskip

\begin{figure*}\vspace{-3mm}
$$\includegraphics[width=13cm]{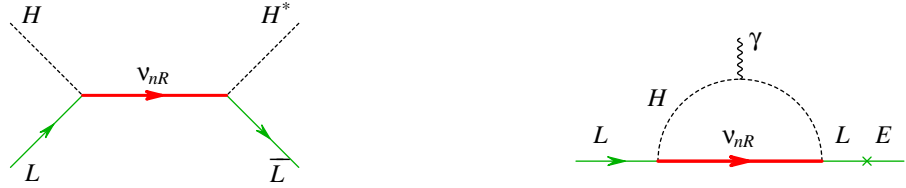}$$\vspace{-1.5cm}
\begin{center}
\caption[X]{\label{fig:virtual} 
Effects generated by virtual exchange of extra dimensional  sterile neutrinos.
At tree level (fig.\fig{virtual}a) they only give the effective operator $(H^\dagger \bar{L}_i)i\ds (HL_j)$.}
\end{center}
\end{figure*}

So far we discussed signals of a single sterile neutrino mixed with $\nu_1$.
The sterile neutrino could mix with a different combination of active neutrinos.
The main qualitative difference is that cosmology becomes the dominant probe
if $\nu_e$ were negligibly involved
in sterile mixing; see~\cite{CMSV} for a precise study.

\section{Neutrinos in extra dimensions}
We now discuss a specific model with an infinite number of sterile neutrinos.
If right-handed neutrinos are light and propagate in large flat extra dimensions,
each of their Kaluza-Klein excitations $\nu_{nR}$
($n=1,2,3,\ldots$)
behaves as a sterile neutrino.
In the minimal model~\cite{Dimo} all their masses $m_n$ and mixings $\theta_n$
are predicted in term of one parameter:
the radius of the extra dimension $R$:
$$m_n \sim \frac{n}{R},\qquad
\theta_n \sim \frac{m_\nu}{m_n}.$$
This kind of physics was much studied a few years ago
as an alternative interpretation of solar and atmospheric anomalies,
but data killed this activity.
Here I try to summarize only what remains clearly alive,
in my opinion.

\subsection*{Supernov\ae}  
Many KK are a serious problem because in a supernova easily encounter many
MSW  level crossings with active neutrinos,
strongly suppressing their fluxes.
One can avoid any level-crossing by demanding that all KK are heavier than
$1/R\circa{>}\sqrt{A E_\nu}\sim\,\hbox{keV}$,
where $A\sim\eV$ and $E_\nu\sim10\MeV$
 are the typical matter potential and neutrino energy.
 This gives $R\circa{<}\AA$.
 Data tolerate slightly higher values of $R$:
some fraction of active neutrinos gets of course lost into the extra dimension,
but there is something more interesting. 
The chemical composition of the SN
gets modified in a way which tends to block $\nu_{nR}$ emission (e.g.\ such that $A_e=0$), resulting
in non standard flavor ratios among fluxes of active $\nu$ and $\bar\nu$~\cite{CR}.
(Nobody computed the prediction of the minimal model~\cite{Dimo} here considered).

\subsection*{Virtual effects}  
The second probe is related to virtual effects i.e.\ by processes in which sterile neutrinos
appear as intermediate particles of Feynman diagrams, see fig.\fig{virtual}.
The interesting observation is that at tree level sterile neutrinos affect only neutrinos
(and Higgs bosons), see fig.\fig{virtual}a.
Charged leptons are affected only by higher order diagrams, see fig.\fig{virtual}b.

In this way it is possible to get effects in neutrinos,
such as flavour transitions at zero baseline
induced by anomalous neutrino couplings (fig.\fig{virtual}a)
$$
P(\nu_i\to \nu_j; L\approx 0) \sim |\epsilon_{ij}^2|,$$
compatibly with existing constraints
from charged leptons (fig.\fig{virtual}b)~\cite{Virtual}
$${\rm BR}(\ell_i\to \ell_j \gamma) \sim |e\epsilon_{ij}/4\pi|^2.$$
In the above equations $i\neq j= \{e,\mu,\tau\}$.
The effects are mediated by sterile neutrinos
with TeV-scale lepton-number conserving masses:
two concrete realizations are pseudo-Dirac sterile neutrinos in four dimensions
and the extra-dimensional model of~\cite{Dimo}.
This model cannot predict the overall value of $\epsilon_{ij}$
(presumably of order $(M_Z/\TeV)^2\sim 10^{-\rm few}$)
but can predict their flavour structure:
 $\epsilon_{ij}\propto ( m_\nu m_\nu^\dagger)_{ij}$~\cite{Virtual}.

\section{LSND}
The sterile interpretation of the LSND anomaly~\cite{LSND} has been disfavored
by a combination of other neutrino experiments~\cite{SN3+1,LSND3+1}
and by standard cosmology~\cite{3+1cosmo}.
Nevertheless it is not excluded, and should also
give $\nu_\mu$ disappearance at a detectable level.
It is difficult to invent alternative plausible speculations cleanly compatible with all data.
For a clear answer see~\cite{MiniBoone}.

\section{Conclusion}
Sterile neutrinos could manifest in the many different ways discussed above.
Solar, supernova, cosmological and other probes have some overlap,
which can  be read from fig.\fig{S1}.
None of the signals we considered is new.
All can (and hopefully will) be searched as a byproduct of future 
neutrino and cosmological experiments.

We emphasize that two probes which seem particularly sensitive
need improvements of `traditional' experiments:
detection of sub-MeV solar neutrinos,
determinations of the primordial Helium-4 abundancy.

\paragraph{Acknowledgments}
I thank M. Cirelli, G. Marandella, A. Notari, F. Vissani, J. Lesgourgues, R. Rattazzi
and the organizers of the Neutrino 2004 conference (`$\nu\,04$').


\begin{thebibliography}{9}
\bibitem{sun} 
An oscillation pattern is now visible in the KamLAND results presented by G. Gratta at $\nu\,04$.
Assuming oscillations,  global fits
find a non-trivial consistency among KamLAND and solar data,
as  reviewed by S. Goswami at  $\nu\,04$.
\bibitem{atm}
An oscillation pattern is now visible in the K2K results presented by T. Nakaya at $\nu\,04$.
The sub-sample of SK data with higher resolution in $L$ and $E_\nu$ shows a hint of a oscillation dip,
as presented  by E. Kearns at $\nu\,04$.
\end{thebibliography}
\end{document}